\documentclass[doublecol,figures]{epl2}
\usepackage{amsmath}
\usepackage{amssymb}
\usepackage{amsfonts}
\usepackage{units}
\usepackage{ae}
\title{High-temperature excess current and quantum suppression of electronic backscattering.}
\shorttitle{High-temperature excess current.}
\author{G. Sonne\inst{1}\thanks{E-mail: \email{gustav.sonne@physics.gu.se}} \and L. Y. Gorelik\inst{2} \and R. I. Shekhter\inst{1} \and M. Jonson\inst{1,3}}
\shortauthor{G. Sonne \etal}
\institute{ 
\inst{1} University of Gothenburg, Department of Physics - SE-412 96 G\"oteborg, Sweden, EU\\
\inst{2} Chalmers University of Technology, Department of Applied Physics - SE-412 96 G\"oteborg, Sweden, EU\\
\inst{3} Heriot-Watt University, School of Engineering and Physical Sciences - Edinburgh EH14 4AS, Scotland, UK, EU
}

\abstract{
We consider the electronic current through a one-dimensional conductor
in the ballistic transport regime and show that the quantum
oscillations of a weakly pinned single scattering target results in a
temperature- and bias-voltage independent excess current at large bias
voltages. This is a genuine quantum effect on transport that derives
from an exponential reduction of electron backscattering in the
elastic channel due to quantum delocalisation of the scatterer and
from a suppression of low-energy electron backscattering in the
inelastic channels caused by the Pauli exclusion principle. We show
that both the mass of the target and the frequency of its quantum
vibrations can be measured by studying the differential conductance
and the excess current. We apply our analysis to the particular case
of a weakly pinned C$_{60}$ molecule encapsulated by a single-wall
carbon nanotube and find that the discussed phenomena are
experimentally observable.
}
\begin{document}
\maketitle
Geometrical confinement is known to alter the interactions in a wide
range of interesting systems. Restricting molecules to the hollow
interior of a single-wall carbon nanotube (SWNT), e.g. has proven to
be an enabling technology for novel chemistry where reactions not
allowed in free space can be studied~\cite{Iijima}. Electron transport
through such encapsulated molecular systems has also been studied to
some extent, both experimentally~\cite{Hornbaker,Utko} and
theoretically~\cite{Kane,Kondo,Kim}, mainly for the the case of the
"carbon peapod", a chain of fullerenes inside a SWNT. Even so, little
is known about the effect of encapsulated neutral molecules on
electron transport along the tube, which is the question addressed in
this Letter.

We focus on the fact that weakly pinned encapsulated molecules can
move rather easily along the nanotube and examine the effect on
electron transport along a SWNT due to scattering from such a
target. To model this we assume the motion of the molecule to be
confined by a shallow harmonic potential well, with the amplitude of its
quantum fluctuations, $X_0$, comparable to the Fermi wavelength,
$\lambda_F=2\pi/k_F$, of the electrons
($k_F\sim$\unit[8.5]{nm$^{-1}$} is the Fermi wave
vector for electrons in a metallic SWNT~\cite{Ouyang}). Below we will
show that the transport properties of the system will be significantly
modified by an exponential reduction of electron backscattering in the
elastic channel due to quantum delocalisation of the scatterer in
conjunction with a suppression of low-energy electron backscattering
in the inelastic channels caused by the Pauli exclusion principle. In
particular we predict a measurable temperature- and bias-voltage
independent excess current at large bias voltages, where the vibration
energy quantum of the molecule sets the energy scale for the bias
voltage.
%
%
%--------picture--------
\begin{figure}[h]
\onefigure[width=0.4\textwidth]{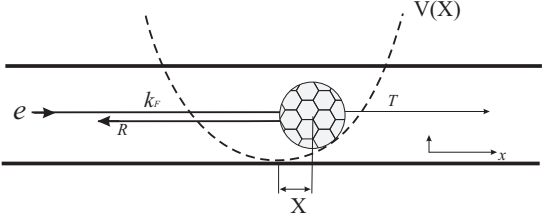}
\caption{Schematic view of the scattering of electrons, $e$, with wave
vector $k_F$ from a movable scattering target in a 
one-dimensional channel. $R$ and $T$ are the reflection and
transmission coefficients respectively and the scattering molecule is
confined by the shallow potential (dashed line) $V(X)=M\omega^2X^2/2$
where $X$ is the deviation of the scatterer from its
equilibrium position.}
\label{picture}
\end{figure}

In order to develop a theory of electronic transport through a SWNT
with an enclosed movable scatterer (sketched in fig.~\ref{picture}) we
describe our system by the Hamiltonian
%
%---first Hamiltonian
\begin{equation}
\label{Hamiltonian1}
\hat{H}=\hat{H}_{el}+\hat{H}_{osc}+\hat{H}_{int}\,.
\end{equation}
%--------
%
Here the first term, 
%
%-----electrons------
\begin{equation}
\hat{H}_{el}=-\frac{\hbar^2}{2m}\int\Psi^{\dag}(x)\frac{\partial^2}{\partial x^2}\Psi(x)\,\upd x \,,
\end{equation}
%----------
%
describes electrons on the nanotube which we treat
as a one-dimensional electronic wire \footnote{Our analysis shows that
the details of the SWNT electronic spectrum do not affect the
phenomena discussed here.}. The $x$-axis is parallel to the
longitudinal direction of the tube and $\Psi^{\dag}(x)$
$\left(\Psi(x)\right)$ creates (annihilates) electrons.
The second term,
%
%-----oscillator
\begin{equation}
\hat{H}_{osc}=\frac{\hat{P}^2}{2M}+\frac{1}{2}M\omega^2\hat{X}^2 \,,
\end{equation}
%-----------
%
models an encapsulated molecule of mass $M$ as a harmonic
oscillator whose zero-point oscillation amplitude
$X_0=\sqrt{\hbar/M\omega}$ may be large if its characteristic
frequency $\omega$ is small; $\hat{P}$ $(\hat{X})$ is the
molecule's momentum (position) operator.

In order to describe the effect of the strong electron-vibron
coupling corresponding to $k_FX_0\sim 1$, one needs to go beyond
the Fr\"ohlich approximation and allow the electron-vibron
interaction Hamiltonian, the third term in (\ref{Hamiltonian1}),
to be strongly nonlinear in the mechanical displacement of the
molecule;
%
%------original interaction
\begin{equation}\label{Hamiltonian_int}
\hat{H}_{int}=\int\Psi^{\dag}(x)\Psi(x)U(x-X)\,\upd x \,.
\end{equation}
%--------
%
Here $U(x-X)$ is the effective potential describing the interaction
between the molecule and the electrons. Since the range of interaction,
$a$, is shorter than the electronic wavelength, $\lambda\sim 1$nm, we
take $U(x-X)=U_0a\delta(x-X)$ where $U_0$ is the effective strength of
the interaction. We will consider small bias voltages,
$V<<\epsilon_F/e$, which means that only electrons close to the Fermi
energy, $\epsilon_F$, contribute to the current. This condition allows
us to separate the electron operators into right- and
left-movers, so that
$\Psi^{\dag}(x)=e^{ik_Fx}\psi_R^{\dag}(x)+e^{-ik_Fx}\psi_L^{\dag}(x)$
where $\psi_{R,L}(x)$ are slowly varying functions of $x$. We neglect
second order derivatives of $\psi_{R,L}(x)$ together with their
variations on the scale of the molecule's vibration amplitude. These
assumptions simplify our Hamiltonian, which can now be expressed as
%
%--------------final Hamiltonian-------------
%\begin{equation}
%\begin{split}
\begin{equation}
\begin{split}
\label{Hamiltonian}
\hat{H}&=\hat{H}_0+\hat{H}_{I}+\hat{H}_{II}+\hat{H}_{osc}\\
\hat{H}_0&=iv_F\hbar\int \bigg(\psi^{\dag}_{R}(x)\frac{\partial}{\partial x}\psi_{R}(x)-\left.\psi^{\dag}_{L}(x) \frac{\partial}{\partial x}\psi_{L}(x)\right)\,\upd x\\
\hat{H}_{I}&=U_0a\bigg(e^{2ik_F\hat{X}}\psi^{\dag}_R(0)\psi_L(0)+\textrm{h.c.}\bigg)\\
\hat{H}_{II}&=U_0a\bigg(\psi^{\dag}_R(0)\psi_R(0)+\psi^{\dag}_L(0)\psi_L(0)\bigg)\,,
\end{split}
\end{equation}
%\end{split}
%\end{equation}
%---------------------
%
where, $v_F=\hbar k_F/m$ is
the Fermi velocity of the electrons. Below we investigate how the
motion of the scattering molecule affects the 1-dimensional 
transport of electrons through our system. To do this we calculate the
current, $I(x_0)$, through the nanotube at some point $x_0$ outside
the scattering region, where
%
%--------current operator
\begin{equation}
\begin{split}
&I(x_0)=Tr\left(\hat{J}(x_0)\hat{\rho}\right)\\
&\hat{J}(x_0)=ev_F\bigg(\psi^{\dag}_R(x_0)\psi_R(x_0)-\psi^{\dag}_L(x_0)\psi_L(x_0)\bigg)\,.
\end{split}
\end{equation}
%----------
%
Here, $\hat{J}(x_0)$ is the current operator at point $x_0$ and
$\hat{\rho}$ is the stationary density matrix satisfying
$[\hat{\rho},\hat{H}]=0$. Considering that the flow of right (left)
moving electrons to the left (right) of the scattering region is
determined by the emission of electrons from the left (right)
reservoirs, kept at the constant chemical potentials $\mu_{l(r)}$, one
finds that the current can be expressed as
%
%-----current-----------
\begin{equation}
\label{current}
I=I_0+\textrm{Im}\frac{4eU_0a}{\hbar}Tr\left(e^{2ik_F\hat{X}}\psi_R^{\dag}(0)\psi_L(0)\hat{\rho}\right)\,.
\end{equation}
%-------------
%
The first term in (\ref{current}), $I_0=G_0V$, is the current without
the scattering center ($G_0=4e^2/h$ is the quantum conductance for a
metallic SWNT) while the second term describes the back-flow current
generated by electronic reflection from the molecule. We calculate the
back-flow current to lowest order in the electron-molecule
interaction, which is assumed to be weak on the energy scale of the
electrons, $U_0<<\hbar v_F/a$. The current is evaluated by expanding
the density matrix in (\ref{current}) to first order in $U_0$,
%
%-------density eq-----
\begin{equation}
\begin{split}
\label{density}
\hat{\rho}=\hat{\rho}_0-\frac{i}{\hbar}\int_{-\infty}^0&\left[\hat{H}_I(t),\hat{\rho}_0\right]\,\upd t\\
\hat{H}_I(t)=e^{i\left(\hat{H}_0+\hat{H}_{osc}\right)t/\hbar}&\hat{H}_Ie^{-i\left(\hat{H}_0+\hat{H}_{osc}\right)t/\hbar} \,,
\end{split}
\end{equation}
%--------
%
where, $\hat{H}_I(t)$ is the time dependent scattering Hamiltonian in
the interaction representation and $\hat{\rho}_0$ is a density matrix
for the non-interacting system. The latter takes the proper boundary
conditions into account, i.e.  right/left-moving electrons are in
thermal equilibrium with the left/right lead with chemical potential
$\mu_{l(r)}=\epsilon_F\pm eV/2$ and the harmonic oscillator is in thermal
equilibrium with the environment. Note that in this approximation
contributions from $\hat{H}_{II}$ do not affect the current.

Using (\ref{current}) together with (\ref{density}) allows us to
calculate the current by evaluating the trace and integrating out the
time variable,
%
%-----current----------
\begin{equation}
\begin{split}
\label{current:eq}
I=G_0V-\frac{G_0R}{e}&\sum_{n=0}^{\infty}P(n)\times\\
\sum_{\Delta=-n}^{\infty}&\vert\langle n\vert e^{i\sqrt{\alpha}(\hat{b}+\hat{b}^{\dag})}\vert n+\Delta\rangle\vert^2F(\Delta\hbar\omega,eV,\beta)\\
F(\Delta\hbar\omega,eV,\beta)=&\int \upd\epsilon\bigg[f_R(\epsilon)\left(1-f_L(\epsilon-\hbar\omega\Delta)\right)-\\
&f_L(\epsilon)\left(1-f_R(\epsilon-\hbar\omega\Delta)\right)\bigg]\\
=&\left(\frac{\hbar\omega\Delta-eV}{e^{\beta(\hbar\omega\Delta-eV)}-1}-\frac{\hbar\omega\Delta+eV}{e^{\beta(\hbar\omega\Delta+eV)}-1}\right)\,.
\end{split}
\end{equation}
%--------
%
In (\ref{current:eq}), $R=\left(U_0a/\hbar v_F\right)^2$ is the (small) reflection
coefficient for electrons,
$P(n)=\left(1-e^{-\beta\hbar\omega}\right)e^{-n\beta\hbar\omega}$
[where $\beta=1/k_BT$] is the probability that the oscillator is in
state $n$ with energy $n\hbar\omega$, and
$f_{R,L}(\epsilon)=(1+e^{\beta(\epsilon-\mu_{l,r})})^{-1}$ are Fermi
distribution functions for right and left moving electrons
respectively. The electron-oscillator coupling is described by the
matrix element $\langle n\vert
e^{i\sqrt{\alpha}(\hat{b}+\hat{b}^{\dag})}\vert n+\Delta\rangle$
between the oscillator states $\vert n\rangle$ and $\vert
n+\Delta\rangle$ where the (dimensionless) parameter $\alpha=\vert
\sqrt{2}k_FX_0\vert^2$ measures the strength of the coupling and
$\hat{b}^{\dag}\,[\hat{b}]$ is a boson creation [annihilation]
operator.

Clearly, the function $F(\Delta\hbar\omega,eV,\beta)$ in
(\ref{current:eq}) contains the effects of reflections of
right/left-moving electrons with the arguments in the respective Fermi
distributions dictating the allowed transitions of the oscillator and
the combination of the Fermi distributions ensuring that the Pauli exclusion
principle is not violated. By evaluating (\ref{current:eq}) one finds
the current through the system as a function of voltage and
temperature. 

First and foremost we find that at high voltages, $eV>eV_0$
($eV_0\propto {\rm max}\{k_BT, \hbar\omega, \Delta\hbar\omega\}$)
there is an excess current, $\delta I$, compared to the ``static''
current $I_{st}=G_0(1-R)V$ that would flow if the scatterer
was immobile. This is because with a mobile scatterer some of the
inelastic reflection channels are blocked due to Pauli principle
restrictions and this reduces the back-flow current. It is
particularly interesting to note that in the limit of very high
voltages, $V\gg V_0$, the excess current is temperature and voltage
independent,
%
%-----limit current----------
\begin{equation}
\label{limitcurrent:eq} \lim_{V\gg V_0}\delta I=\lim_{V\gg
V_0}I-I_{st}=\frac{4G_0 R}{e}\left(\frac{\hbar^2 k_F^2}{2M}\right)\,.
\end{equation}
%------------
%
This is a remarkable result, not only in that it predicts
the high voltage excess current --- a truly quantum-mechanical
phenomenon --- to be independent of temperature, but also in that it
should be feasible to probe experimentally.
%
%-------current fig------------
\begin{figure}
\onefigure[width=0.4\textwidth]{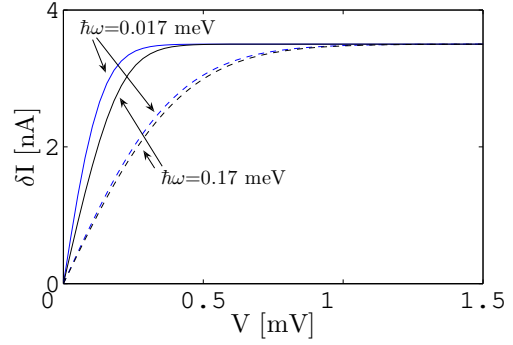}
\caption{Excess current as a function of voltage for a SWNT with an
enclosed C$_{60}$ for two different vibration energies
$\hbar\omega$. The high voltage limit of the excess current is seen to
be independent of oscillator frequency, voltage and temperature and
scales $\propto \hbar^2 k_F^2/(2M)$. Here, $R=0.3$, $k_F=$\unit[8.5]{nm$^{-1}$}
\protect\cite{Ouyang} with $T=$\unit[0.5]{K} (solid) and $T=$\unit[1.5]{K}
(dashed).}
\label{current:fig}
\end{figure}
%-------
%

Also interesting to note is that the amount of excess current scales
as $\hbar^2k_F^2/(2M)$; hence it is inversely proportional to the mass
of the target and does not depend on the curvature of the harmonic
potential. Even more, one can prove that the result in
(\ref{limitcurrent:eq}) is quite general and valid for any confining
potential.

One can only speculate about the possible value of $\hbar\omega$
for the longitudinal oscillatory fullerene motion along the
nanotube. Here, we estimate its value to be one to two orders of
magnitude lower than the \unit[5]{meV} reported by Park \textit{et al.}
for a C$_{60}$ molecule bound to a gold electrode by van der Waals
forces \cite{Park}. The excess current is shown in
fig.~\ref{current:fig} for two different temperatures and two
different confining potentials.

\begin{figure}[h]
\onefigure[width=0.35\textwidth]{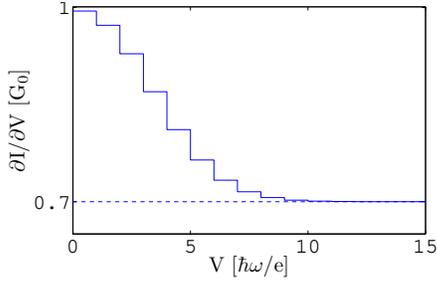}
\caption{Low temperature differential conductance for the same
system as in fig.~\ref{current:fig} shown as a function of voltage
for $\alpha=5$. The dashed line shows the limiting value $\partial
I/\partial V=G_0\left(1-R\right)$. Here, $R=0.3$.}
\label{fig:lowTdiff}
\end{figure}

From (\ref{current:eq}) one can also calculate the low temperature
differential conductance. By using the restrictions imposed by the
allowed transitions between oscillator states and by noting that
as $T\rightarrow 0$, $P(n)=\delta_{n,0}$, one finds that
%
%-----differential cond----------
\begin{equation}
\label{limitdiff:eq} \lim_{T\rightarrow 0}\frac{\partial
I}{\partial
V}=G_0\left(1-R\sum_{\Delta=0}^{\left[\frac{eV}{\hbar\omega}\right]}\frac{e^{-\alpha}\alpha^{\Delta}}{\Delta!}\right)\,.
\end{equation}
%---------------------
%
Here, $\left[eV/\hbar\omega\right]$ is the integer part of the
ratio between the applied bias voltage and the characteristic
energy scale of the oscillator, which dictates the number of open
scattering channels. Thus, as the applied voltage across the
system is increased the low temperature differential conductance
is seen to approach the expected result for a immobile scatterer,
$G_0\left(1-R\right)$, as shown in fig.~\ref{fig:lowTdiff}. This
is understandable since a larger bias voltage will increase the
number of possible final oscillator states, making the motion of
the fullerene less important on the energy scale of the electrons
and the result from a static barrier due to Landauer
\cite{Landauer} is recovered.
%
%------diff condu fig--------
%----------
%

\begin{figure}
\onefigure[width=0.49\textwidth]{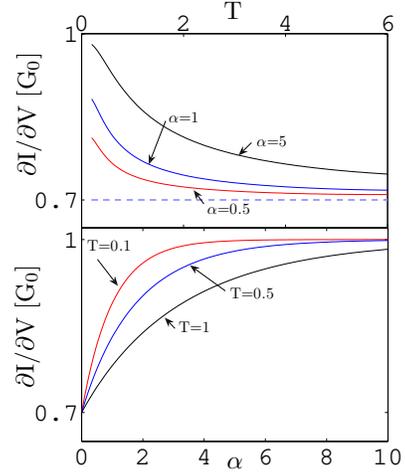}
\caption{Linear conductance for the same system as in
fig.~\ref{current:fig} where $R=0.3$, shown (top) as a function of
temperature $T$ in units of $[\hbar\omega/k_B]$ and (bottom) as a
function of electron-vibron coupling parameter $\alpha$.}
\label{fig:condu}
\end{figure}

The linear conductance of the system refers to the limit of zero
bias voltage and generally has to be evaluated numerically.
However, its low and high temperature asymptotic limits can be
found using the completeness of the set of vibronic states,
$\sum_{\Delta=0}^{\infty}\vert \langle n\vert
e^{i\sqrt{\alpha}(\hat{b}+\hat{b}^{\dag})} \vert \Delta\rangle\vert^2=1$.
These limiting results, which are similar to those found by
Shekhter \textit{et al.} \cite{Shekhter} for the case of a
suspended nanotube in a transverse magnetic field, can be
expressed as
%
%-------------
\begin{equation}
\label{limit:G}
%\lim_{V\rightarrow0}
\frac{\partial I}{\partial V}
=G_0\left\{ \begin{array}{ll}
\left(1-Re^{-\alpha}\right) & \beta\hbar\omega\gg 1\\
\left(1-R\left[1-\frac{\alpha\hbar\omega}{3k_BT}\right]\right) & \beta\hbar\omega\ll 1\,.
\end{array} \right.
\end{equation}
In the low temperature limit, $\beta\hbar\omega \gg 1$, only the
ground state elastic channel is allowed and the conductance goes as
the zero-bias limit of (\ref{limitdiff:eq}). In the high temperature
limit, $\beta\hbar\omega <<1$, many inelastic channels for
back-scattering are available, which reduces the excess conductance of
the system as compared to a static barrier case.  These asymptotes are
clearly visible in the numerical solutions for the linear conductance,
fig.~\ref{fig:condu}. Also of interest is the low temperature
conductance as a function of $\alpha$, also shown in
fig.~\ref{fig:condu}. We propose that this region could be studied
experimentally by tuning the Fermi wavelength, hence $\alpha$, using a
gate voltage.
%
%-----conductance figure-------
%----------
%

It is interesting to note that in all the above cases the energy
scale is set by the frequency of the fullerene oscillator, i.e.
$\hbar\omega$ separates out low and high temperature effects in
the conductance of our carbon nanotube-system. For
$\hbar\omega\sim$\unit[0.1]{meV} the cross-over energy would correspond
to a temperature of order $\sim$\unit[1]{K}.

It is also of interest to compare our result with that found by Krive
\textit{et al.} \cite{Krive} for resonant electron tunneling through
carbon peapods. In agreement with preliminary experimental results
\cite{Utko2} they found an anomalous $T^{-1/2}$ temperature dependence
of the conductance, which they attribute to a polaron-assisted
tunneling. Their model is, however, specific to resonant tunneling and
is complementary to ours, which applies to the ballistic transport
regime.

To conclude, our analysis has shown that in the weak scattering limit,
confinement of a molecule inside a carbon nanotube leads to quantum
corrections to both the current and the conductance through the
system. In particular we have shown that due to the quantum
fluctuations of the scattering target, a temperature and voltage
independent excess current is predicted at high bias. We find that the
magnitude of the excess current in the high voltage limit is itself
not a function of the confining potential but scales inversely with the
mass of the target, thus making this an observable quantity for any
confined scattering target for which the onset of the excess current
is dictated by the confining potential. An extension of our
perturbative approach to the case of many encapsulated molecules should be
possible for low enough molecule concentrations, in which case they
can be treated independently, allowing us to sum up their individual
contributions.

This work was supported in part by the Swedish VR and SSF and by
the EC through project FP6-IST-003673 CANEL. The views expressed
in this publication are those of the authors and not necessarily
those of the EC.
%
%--------bibliography
\bibliography{/chalmers/users/sonneg/Desktop/Artiklar/referencespeapod}
\bibliographystyle{/chalmers/users/sonneg/Desktop/PeapodProject/Europhys/eplbib}

\end{document}